\documentclass[aps,prb,twocolumn,nowpacs,superscriptaddress,groupedaddress,amsmath,amssymb]{revtex4}
\usepackage {graphicx,epsfig,graphics,color}

\makeatletter

\newcommand{\Rmnum}[1]{\expandafter\@slowromancap\romannumeral #1@}
\makeatletter
\begin{document}

%\linenumbers
%\def\linenumberfont{\normalfont\small\sffamily}
%\setpagewiselinenumbers
%\modulolinenumbers[5]

\title{Large magnetic penetration
depth and thermal fluctuations in a
Ca$_{10}$(Pt$_{3}$As$_{8}$)[(Fe$_{1-x}$Pt$_{x}$)$_{2}$As$_{2}$]$_{5}$
(x=0.097) single crystal}

\author{Jeehoon Kim}
\affiliation{Los Alamos National Laboratory, Los Alamos, NM 87545}
\email[Corresponding author: ]{jeehoon@lanl.gov}
\author{Filip Ronning}
\affiliation{Los Alamos National Laboratory, Los Alamos, NM 87545}
%\email[Electronic mail: ]{fronning@lanl.gov}
\author{N. Haberkorn}
\affiliation{Los Alamos National Laboratory, Los Alamos, NM 87545}
\author{L. Civale}
\affiliation{Los Alamos National Laboratory, Los Alamos, NM 87545}
\author{E. Nazaretski}
\affiliation{Brookhaven National Laboratory, Upton, NY 11973}
\author{Ni Ni}
%\affiliation{Department of Chemistry, Princeton University, Princeton, NJ 08544}
%\author{J. M. Allred}
\affiliation{Department of Chemistry, Princeton University, Princeton, NJ 08544}
\author{R. J. Cava}
\affiliation{Department of Chemistry, Princeton University, Princeton, NJ 08544}
\author{J. D. Thompson}
\affiliation{Los Alamos National Laboratory, Los Alamos, NM 87545}
\author{R. Movshovich}
\affiliation{Los Alamos National Laboratory, Los Alamos, NM 87545}

\date{\today}

\begin{abstract}
We have measured the temperature dependence of the absolute value
of the magnetic penetration depth $\lambda(T)$ in a
Ca$_{10}$(Pt$_{3}$As$_{8}$)[(Fe$_{1-x}$Pt$_{x}$)$_{2}$As$_{2}$]$_{5}$
(x=0.097) single crystal using a low-temperature magnetic force
microscope (MFM).  We obtain $\lambda_{ab}$(0)$\approx$1000 nm via
extrapolating the data to $T = 0$. This large $\lambda$ and
pronounced anisotropy in this system are responsible for large
thermal fluctuations and the presence of a liquid vortex phase in
this low-temperature superconductor with critical temperature of
11 K, consistent with the interpretation of the electrical
transport data. The superconducting parameters obtained from
$\lambda$ and coherence length $\xi$ place this compound in the
extreme type \MakeUppercase{\romannumeral 2} regime. Meissner
responses (via MFM) at different locations across the sample are
similar to each other, indicating good homogeneity of the
superconducting state on a sub-micron scale.
\end{abstract}

%\pacs{07.79.Pk, 07.55.-w, 76.50.+g, 75.70.-i}
\maketitle

Iron-based superconductors offer an opportunity to explore
superconductivity over a very wide range of superconducting
properties, such as critical fields, superfluid densities, and
their anisotropy.\cite{Paglione 2010} Comparing iron-based
superconductors with cuprates provides clues to the mechanism of
high $T_{c}$ superconductivity that determine fundamental
superconducting parameters, such as the gap symmetry\cite{Yin
2009,Chubukov 2009} and the upper critical fields,\cite{Yuan
2009} as well as complex vortex dynamics due to the thermal
fluctuations.\cite{Blatter,Prozorov 2008,Beek 2010} Understanding
the correlation between intrinsic properties and the pinning
mechanism is thus intriguing from both basic and applied points
of view. Recently, superconductivity has been reported in a new
family of highly anisotropic materials;
Ca$_{10}$(Pt$_{n}$As$_{8}$)[(Fe$_{1-x}$Pt$_{x}$)$_{2}$As$_{2}$]$_{5}$
(Ca-Pt-Fe-As) with n=3 (``10-3-8") and n=4 (``10-4-8").\cite{Ni
2011,Kakiya 2011,Lohnert} The 10-3-8 phase has $triclinic$
symmetry and a $T_{c}$ of up to 11 K upon Pt doping; the 10-4-8
phase has tetragonal symmetry with the highest $T_{c}$ of 38 K.
It is worth noting that well-defined Fermi surface sheets with
tetragonal symmetry, similar to other pnictides, are observed in
the 10-3-8 phase in spite of its triclinic
symmetry.\cite{Neupane} Anisotropy of the critical field in the
10-3-8 phase near $T_{c}$, $\gamma_{H_{c2}}(T_{c})\equiv
H^{ab}_{c2}/H^{c}_{c2}\approx 10 $ ( Ref. 8), is much larger than
that of the 122 pnictide compounds,
$\gamma_{H_{c2}}(T_{c})\approx 2$ (Ref. 12), consistent with a
more anisotropic 2D nature of the 10-3-8 system. In contrast to
Ba(Fe$_{1-x}$Co$_x$)$_{2}$As$_{2}$ superconductors, well-studied
by a variety of techniques,\cite{Ni 2008} 10-3-8 shows a broadened
superconducting transition temperature with applied
field,\cite{Ni 2011} consistent with strong thermal fluctuations
of vortices.\cite{Blatter}

In this Rapid Communication we present measurements of the
absolute value of the magnetic penetration depth $\lambda$ in the
10-3-8 compound. We derive the values of several basic
superconducting parameters from our measurements and relate them
to other unusual properties observed in the 10-3-8 compound, such
as a broadened superconducting phase transition. We have
determined the temperature dependence of the absolute value of
$\lambda(T)$ in a 10-3-8 single crystal (x=0.097)\cite{Ni 2011}
with $T_{c}$ $\approx$ 11 K using a Meissner technique employing
magnetic force microscopy (MFM). Our experimental approach for
$\lambda$ measurements is simple, robust, and independent of the
MFM tip model whereas most previous MFM studies provided either
$\delta\lambda$ variations or the absolute value from modeling
the tip magnetization instead of measuring $\lambda$
directly.\cite{Luan 2010,Nazaretski 2009} Recently, the
temperature dependence of $\delta\lambda$ was measured via a
tunnel-diode resonator technique, and it showed an increasing
anisotropy of the superconducting gap as doping decreases from
optimal doping towards the edges of the superconducting
dome.\cite{Cho} Our MFM results show that the superconductivity
is homogeneous, which agrees well with tunnel diode,\cite{Cho}
transport,\cite{Ni 2011} and angle resolved photoemission
spectroscopy (ARPES) data.\cite{Neupane} By extrapolating our
temperature dependent data from 4 K to $T=0$, we obtain
$\lambda_{ab}$(0)$\approx$1000 nm. The short electron mean free
path in this system, compared to the coherence length, suggests
that this system is in the dirty limit, which is partly
responsible for the large $\lambda$ value. Strong thermal
fluctuations inferred from the Ginzburg number are consistent
with a wide superconducting transition under field and the
presence of a vortex-liquid phase in this highly anisotropic
superconductor with relatively {\it low} $T_{c}$.\cite{Ni 2011}

\begin{figure}
\centering
\includegraphics [trim=0 0 0 0cm,clip=true,width=8.5cm] {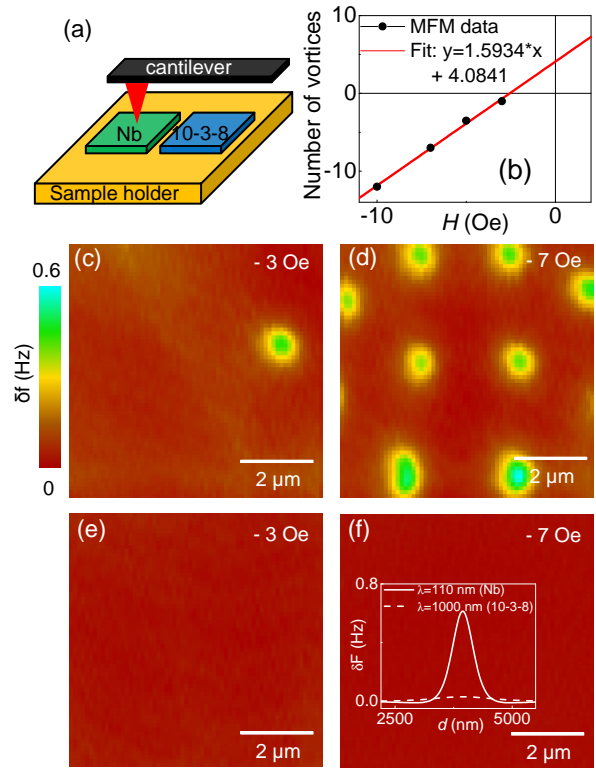}
\caption {\label{f:stray} (Color online) (a) Schematic
illustration of our sample holder with multiple samples. A Nb
thin film (300 nm) and a 10-3-8 single crystal are located next
to each other. (b) The number of vortices as a function of the
magnetic field. The red (light gray) line represents a linear fit
to the experimental data. (c) and (e) MFM images of the Nb sample
and the 10-3-8 sample taken at $T=$4 K in $H=$-3 Oe,
respectively. (d) and (f) The same type of MFM images as in (c)
and (e) but with $H=$-7 Oe. No individual vortices were clearly
resolved in 10-3-8 due to a large $\lambda$ ((e) and (f)) as
opposed to the Nb reference. The color scale bar is applied for
(c)-(f). The inset in panel (f) shows a simulation of the
expected frequency shift for two different $\lambda$ values. The
solid line represents a calculated line profile for Nb and the
dashed curve shows a profile for a 10-3-8 sample based on the
monopole-monopole interaction between the tip and the sample. The
frequency shift of the vortex center in a 10-3-8 sample is around
30 mHz (close to the noise level). Small frequency shift prevents
visualization of vortices.}
\end{figure}

Synthesis of the 10-3-8 system is described elsewhere.\cite{Ni
2011} All measurements described here were performed in a
home-built low-temperature MFM apparatus,\cite{Nazaretski RSI
2009} which allows acquisition of a complete set of MFM data on
several samples with identical MFM tip conditions. With this
apparatus, a Meissner response curve\cite{Luan 2010} is measured
first as a function of the tip-sample separation in the reference
sample (Nb film), and then, the cantilever is moved to the sample
of interest (10-3-8) where its Meissner response as a function of
tip-sample distance is obtained. Direct comparison of these
curves yields the absolute value of $\lambda_{ab}$ in 10-3-8 and
its temperature dependence $\lambda_{ab}(T)$. (We measure
$\lambda_{ab}$ since the shielding currents run in the basal
plane.) The $\lambda$ value of the reference Nb film was verified
by both a different MFM technique and SQUID magnetometry
measurements, as described elsewhere.\cite{Nazaretski 2009} The
vortex imaging, after field-cooling the sample in a field of a few
Oersted to avoid the demagnetization effect of the sample, was
performed in a frequency modulation mode with the tip-lift height
of 400 nm above the sample surface. The zero point of the
tip-sample separation was determined by touching the surface of
the sample; this touchdown of the tip resulted in a substantial
negative frequency shift. The tip-sample separation was measured
based on the calibration of the piezo scanner. We used a high
resolution cobalt-coated Nanosensors cantilever\cite
{Nanosensors} that was polarized along the tip axis in a 30-kOe
magnetic field ($H$). The Meissner experiment was performed under
the conditions of no vortices being present in a 20 $\mu$m
$\times$ 20 $\mu$m field of view, which eliminates possible force
contributions of vortices to the Meissner force. Before vortex
measurements on the 10-3-8 sample, the stray field ($H_{sf}$)
from a superconducting magnet was calibrated by measuring the
number of vortices as a function of field on the Nb reference
(see Fig.~\ref{f:stray}(b)): the Nb reference serves as a
magnetic field sensor. The red line in Fig.~\ref{f:stray}(b) is a
linear fit to the experimental data with a fit function of
$N=(N/H)H+H_{sf}$, where $N$ is a number of vortices. The
obtained slope and $H_{sf}$ from the fit are 1.6 Oe$^{-1}$ and
-2.6 Oe, respectively. The calculated single vortex flux
$\Phi_{exp}$ from the slope of the fit and the area of a vortex
image is $\Phi_{exp}=area\times(N/H)^{-1}=(6~\text{$\mu$m} \times
6~\text{$\mu$m})\times (1.6)^{-1}=22.6$~G$\mu$m$^{2}$, which is
in good agreement with the theoretical value of a single flux
quantum $\Phi_{0}=h/2e=20.7~\text{G$\mu$m$^{2}$}$. The Nb
reference film is homogeneous with a uniform (though irregular)
distribution of vortices. The stray field calibration in panel (b)
in Fig.~\ref{f:stray} was reproducible at 3 different locations of
the Nb reference.

Figs.~\ref{f:stray}(c)-(f) show MFM images obtained at two
different nominal fields of $H=-3$ Oe and $H=-7$ Oe in both the Nb
reference (panels (c) and (d)) and the 10-3-8 single crystal
(panels (e) and (f)). We obtained MFM images at several locations
across the sample's surface separated by hundreds of microns and
observed no vortices. Thermal drift of the MFM system equals a
few nanometers per hour at 4 K. The low drift comes from a rigid
design of the MFM apparatus. More technical details can be found
in Ref. 15. This indicates that the lack of vortices in a 10-3-8
sample is an intrinsic property. The lack of vortices may be due
to a large $\lambda$, leading to a very slow exponential decay of
the vortex profile over a large length scale, and hence a smaller
intensity of the MFM signal in the vortex
center.\cite{vortex-profile} To verify this possibility, we
measured the Meissner response as a function of the tip-sample
separation. An MFM tip experiences a Meissner force because of
the interaction between the tip magnetic moment and a field
generated from the shielding current induced by the tip moment.
The Meissner response force can be expressed as a function of
$\lambda$ and the tip-sample separation $d$,
$F_{Meissner}=A\times f(d+\lambda)$, where $A$ is a prefactor
that reflects the sensor's geometry and the magnetic
moment.\cite{Luan 2010,Auslaender 2009,Straver 2008,Luan
arxiv,Shapoval 2011} The Meissner forces obtained from a Nb
reference and the 10-3-8 sample have the same functional form,
$F^{Nb}(d) = A\times f(d+\lambda_{Nb})$ and $F^{10-3-8}(d) =
A\times f(d+\lambda_{10-3-8})$, respectively. Note that $A$ and
$f$ are the same in both cases when the tip is at the same
condition. As a result, $\lambda_{10-3-8} =
\lambda_{Nb}+\delta\lambda$, where the reference $\lambda_{Nb}=$
110 nm has been previously determined \cite{Nazaretski 2009} and
$\delta\lambda$ is the shift required to overlay the $F(d)$
curves.

\begin{figure}
\centering
\includegraphics [trim=0.5cm 0 0 11cm,clip=true,angle=0,width=8.5cm] {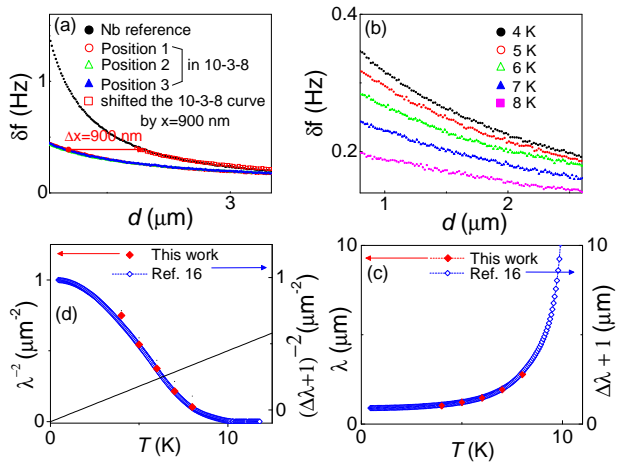}
\caption{\label{f:absolute} (Color online) (a) Meissner curves
obtained from a Nb reference and a 10-3-8 sample with the same
experimental conditions during a single cool-down. The Meissner
curves obtained from 10-3-8 were obtained at three lateral
positions separated by approximately 10 $\mu$m. Their similarity
indicates good homogeneity of the superconducting state on a
sub-micron scale in the sample. By shifting the red empty curve
by 900 nm, the resulting curve (red empty squares) overlays the
Nb reference curve. (b) Temperature dependent Meissner response
curves. Note that the Meissner curves decay more slowly with
increasing tip-sample separation as the temperature increases,
indicating an increase of $\lambda$ with temperature. (c)
Temperature dependent $\lambda$ measurements. Red filled diamonds
are experimental data inferred from curves in
Fig.~\ref{f:absolute}(b). Blue empty diamonds are taken from
tunnel diode resonator measurements.\cite{Cho} We shifted tunnel
diode data along the $y$ axis by 1 $\mu$m, the value obtained
from MFM measurements and overlaid on our data to directly
compare with the MFM data. (d) Superfluid density $\rho_{s}(T)$
calculated from Fig.~\ref{f:absolute}(c). The black line
corresponds to $\rho^{2D}_{s}\equiv
\hbar^{2}d/4k_{B}e^{2}\mu_{0}\lambda^{2}=(2/\pi)T$ in each FeAs
plane.}
\end{figure}

Fig.~\ref{f:absolute}(a) shows Meissner curves obtained from the
Nb film and the 10-3-8 single crystal. The Meissner curves in the
10-3-8 sample were obtained at three lateral positions separated
by approximately 10 $\mu$m: the uniformity of the curves
indicates that $\lambda$ is homogeneous on a sub-micron scale. As
opposed to the Nb reference, the Meissner curve for the 10-3-8
crystal decays slowly as the tip-sample separation increases,
indicating the $\lambda$ for the 10-3-8 compound is larger than
that of Nb. The absolute value of $\lambda$ in the 10-3-8 sample
is obtained by offsetting the 10-3-8 Meissner curve to overlay the
response in the Nb reference. The offset yields $\delta\lambda$ of
900 nm, as shown by the arrow in Fig.~\ref{f:absolute}(a), and
therefore $\lambda_{10-3-8}=1000\pm 100$ nm. The
temperature-dependent penetration depth, shown in
Fig.~\ref{f:absolute}(c), was obtained from Meissner curves
measured at different temperatures, as shown in
Fig.~\ref{f:absolute}(b). Fig.~\ref{f:absolute}(c) displays the
absolute values of of $\lambda(T)$ obtained using MFM (red filled
diamonds) and using tunnel diode measurements\cite{Cho} (blue
empty diamonds). To compare the two data sets, the tunnel diode
data ($\delta\lambda(T)$) are shifted along the $y$ axis by 1
$\mu$m, the absolute value of $\lambda(0)$ obtained from MFM
measurements. These data sets demonstrate consistency, proving
the validity of the MFM approach. We also measured the absolute
value of $\lambda$ in a 10-3-8 sample with a different doping
level ($x=0.042$; $T_{c}\approx$ 10~K) and obtained (a doping
dependence) $\lambda_{ab}(0)\approx 1200\pm$100~nm.

The large $\lambda$ measured here can be due to either impurity
scattering or an intrinsically small superfluid density. To
evaluate the contribution from impurities, we first estimate the
electronic mean free path using the Drude model:
$l=\frac{1}{2\pi}\frac{R_{K}k_{F}}{n\rho}$, where $R_{K}$ is the
von Klitzing constant ($R_{K}=h/e^{2}\approx 25813$~$\Omega$),
$k_{F}$ is the Fermi wave vector, $n$ is the charge carrier
density, and $\rho$ is the resistivity. We obtain $l\approx$1.5
nm, using $n$ and $\rho$ obtained via transport
measurements\cite{Ni 2011} and $k_{F} \approx 0.3\pi /a$ from
ARPES\cite{Lohnert}. The mean free path ($l=1.5$~nm) is shorter
than the coherence length ($\xi=5$~nm) obtained from
transport\cite{Ni 2011} ($l < \xi$), and indicates that the
system is in the dirty limit, partially explaining the large
$\lambda$ value. In the dirty limit the effective penetration
depth is
$\lambda_{eff}(0)=\lambda_{clean}(0)(1+\xi_{0}(0)/l)^{1/2}$ using
the local approximation and the effective coherence length is
$\xi_{eff}=\xi_{clean}(0)/(1+\xi_{0}(0)/l)^{1/2}$.\cite{Tinkham,Kim}
For $\xi_{0}=$19.7~nm~(obtained from the equation of $\xi_{eff}$),
$\lambda_{eff}=$1000~nm, and $l=$1.5~nm, $\lambda_{clean}$ in the
clean limit is approximately 270 nm.

A large $\lambda$ also can be due to a small superfluid density.
The London penetration depth
$\lambda_{L}=\sqrt{\frac{m^{\star}}{\mu_{0}ne^{2}}}$, where
$m^{\star}$ is an effective electron mass, $\mu_{0}$ is the vacuum
permeability, $n$ is the charge carrier density, and $e$ is an
electron charge. ARPES E(k) data allow us to calculate the
effective mass of the charge carriers using the expression
$1/m^{\star}=\frac{1}{\hbar^{2}}(d^{2}E/dk^{2})_{E_{f}}$. We
obtain $m^{\star}\approx$7.3$m_{e}$, where $m_{e}$ is a bare
electron mass\cite{Lohnert}. This $m^{\star}$ and $n\approx
0.74\times 10^{27}$~m$^{-3}$, obtained from transport
measurements,\cite{Ni 2011} results in $\lambda_{L}\approx 530$
nm. This calculated $\lambda_{L}$ corresponds to an intrinsic
$\lambda$, because the experimental band dispersion and carrier
density do not depend on disorder. Therefore, we can directly
compare $\lambda_{clean}\approx 270$~nm with $\lambda_{L}\approx
530$~nm from ARPES and transport. The discrepancy is likely a
result of the carrier density being obtained using a single band
approximation; whereas, ARPES and theoretical electron band
calculations show a multiband character of the Fermi surface.

\begin{figure}
\centering
\includegraphics [trim=1cm 0 0 2cm,clip=true,angle=0,width=9.5cm] {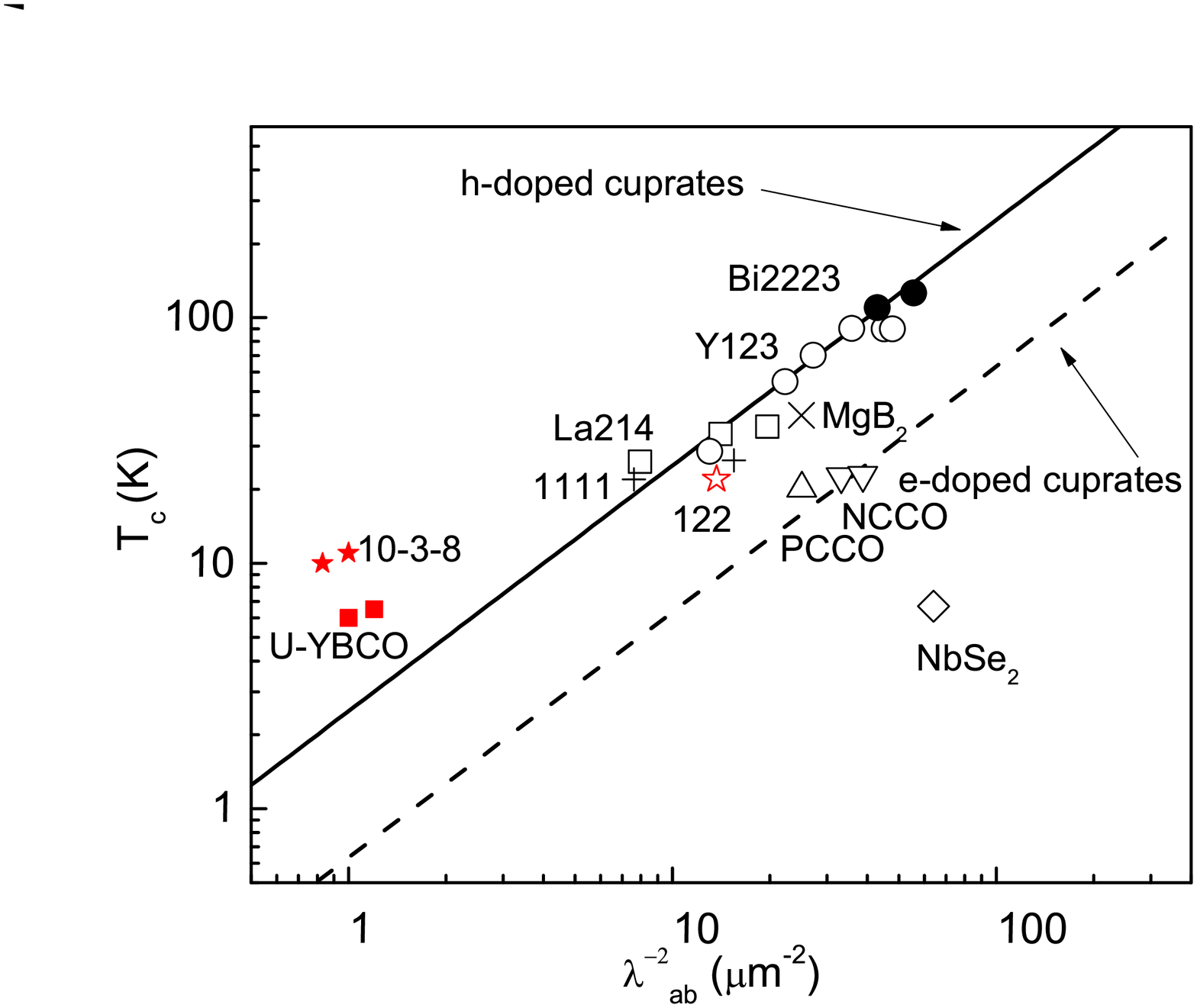}
\caption{\label{f:uemura} (Color online) Uemura plot for high
$T_{c}$ cuprates. The data for the cuprates and points for the
1111 system are taken from Ref. 27. The data for MgB$_2$ and
NbSe$_2$ are taken from Ref. 28 and Ref. 29, respectively. The red
filled squares show highly underdoped YBCO superconductors taken
from Ref. 30. The red filled stars represent the data for 10-3-8
obtained in this work.}
\end{figure}

The small superfluid density, reflected in the large measured
penetration depth, indicates a weak phase stiffness of the
superconducting order parameter and suggests that phase
fluctuations may be important in 10-3-8. We add in
Fig.~\ref{f:uemura} the values of $\lambda$ for 10-3-8 as red
filled stars to the Uemura plot,\cite{Uemura 1989} which shows the
scaling between $\lambda^{-2}(0)\propto n_{s}/m^{\star}$ and
$T_{c}$ in unconventional superconductors. We see that the phase
stiffness relative to $T_c$ is weaker than in other Fe-based
superconductors as well as the cuprates, even weaker than that in
highly underdoped YBCO.\cite{Luetkens 2008,Broun 2007,Zuev 2005}
This may reflect a ``Swiss cheese'' like response of the system to
impurities, indicating nanoscale inhomogeneity is likely present
in 10-3-8.\cite{Franz,Das} Our measurement of $\lambda$ is not
sensitive to heterogeneity on this scale. The relatively large
anisotropy of the upper critical field further suggests that
10-3-8 may be the most quasi-2D material of all Fe-based
superconductors,\cite{Ni 2011} thus bearing more resemblance to
the cuprate superconductors. Fig.~\ref{f:absolute}(d) shows the
measured superfluid density $\rho_s(T)$ as well as prediction of
the Kosterlitz-Thouless-Berezinsky (KTB) theory of
vortex-unbinding that should be applicable to a highly anisotropic
2D superconductor.\cite{BTK} $\rho_s(T)$ passes smoothly through
the KTB line, indicating that superconductivity of the 10-3-8
phase is still of a 3-D character.

We now discuss the effects of the large $\lambda$ on other
superconducting properties. The $H_{c2}$ values obtained from the
Werthamer-Helfand-Hohenberg theory\cite{Werthamer 1966} are
$H_{c2}^{\parallel ab}$ = 55 T and $H_{c2}^{\perp ab}$ = 13 T,
and the corresponding $\xi$ values are $\xi_{\parallel ab}(0)$ =
5 nm and $\xi_{\perp ab}(0)$ = 1.2 nm.\cite{Ni 2011} We calculate
$\kappa=\lambda_{ab}/\xi_{ab}\approx$ 200, using
$\lambda_{ab}\approx 1000$ nm and $\xi_{ab}\approx  5$ nm, the
thermodynamic critical field
$H_{c}=\Phi_{0}/2\sqrt{2}\pi\lambda(0)\xi(0)\approx 500$ Oe, and
the depairing current
$J_{d}=\Phi_{0}/3\sqrt{3}\pi\mu_{0}\lambda^{2}\xi\approx 2$
MAcm$^{2}$. These values indicate that the 10-3-8 compound is an
extreme type \MakeUppercase{\romannumeral 2} superconductor.
Anisotropy of the critical field in this compound $\gamma_{H_{c2}}
= H_{c2}^{\parallel ab} /H_{c2}^{\perp ab}$ shows a strong
temperature dependence, ranging from 10 near $T_{c}$ to 5 at
0.9$T_{c}$.\cite{Ni 2011} The resistive signature of the
superconducting transition broadens with increasing magnetic
field, indicating the presence of strong magnetic
fluctuations.\cite{Ni 2011} The fundamental parameter governing
the strength of the thermal fluctuations is the Ginzburg number
$Gi$, $Gi=[T_{c}\gamma/H^{2}_{c}(0)\xi^{3}(0)]^{2}/2$, where
$H_{c}$ is the thermodynamical critical field and $\gamma$ is the
anisotropy parameter.\cite{Blatter} Using $T_c\approx 11$ K,
$\gamma\approx10$, $\xi = 5$ nm, and $\lambda\approx 1000$ nm, we
obtain $Gi\approx 0.16$. The theoretical width of the transition,
$\Delta T_c\geq Gi\cdot T_{c}$, is approximately 1.8 K, consistent
with the experimental value of $\Delta T_{c}\approx 2$ K;
although, we can not rule out that the rounding is also partially
a result of nanoscale spatial inhomogeneity of T$_{c}$.\cite{Ni
2011} The $Gi$ in the 10-3-8 compound is larger than that in YBCO
($Gi=$0.01) and BiSCCO ($Gi=$0.1).\cite{Blatter} The broadening
of the superconducting transition with increasing magnetic field
is consistent with the presence of a vortex-liquid phase, similar
to cuprates.\cite{Blatter}

In conclusion, we measured the absolute value of $\lambda$ in a
single crystal of
Ca$_{10}$(Pt$_{3}$As$_{8}$)[(Fe$_{1-x}$Pt$_{x}$)$_{2}$As$_{2}$]$_{5}$
(x=0.097) using a Meissner technique in a low temperature MFM
apparatus. Similar Meissner responses in different regions of the
sample indicate that the superconductivity is homogeneous on a
scale of $\lambda$. We obtain the value of $\lambda(0)$ in our
sample of approximately 1000 nm. The clean limit penetration
depth is calculated to be 270 nm based on an impurity scattering
model. The large Ginzburg number ($Gi\approx0.16$) agrees well
with the previously reported data that show a broad
superconducting transition and a signature of a vortex liquid
phase in this highly anisotropic {\it low} $T_{c} = 11$ K
superconductor.

We acknowledge valuable discussions with M. Graf. Work at Los
Alamos was supported by the US Department of Energy, Office of
Basic Energy Sciences, Division of Materials Sciences and
Engineering. Work at Brookhaven was supported by the US
Department of Energy under Contract No. DE-AC02-98CH10886. Work
at Princeton was supported by the AFOSR MURI on
superconductivity, grant FA9550-09-1-0593. We thank K. Cho, M.
Tanatar, and R. Prozorov for supplying data from Ref. 16. N. H. is
member of CONICET, Argentina.


\begin{thebibliography}{99}
\bibitem{vortex-profile} The maximum force gradient at the
center of a vortex, max($\partial f/\partial z$) (MFM is
sensitive to a force gradient), is proportional to
$(z+\lambda_{ab})^{-1/3}$ for a monopole-monopole model of the
tip-vortex interaction. In this model, the larger value of
$\lambda$ results in a smaller force gradient at the center of a
vortex.
\end{thebibliography}
\end{document}